\documentclass[floatfix, noshowpacs, preprintnumbers, twocolumn, amsmath, amssymb, aps, superscriptaddress, prb]{revtex4}

\pdfoutput=1

\usepackage{graphicx, xcolor}
\usepackage[caption=false]{subfig}
\usepackage{soul}
\usepackage{color}
\usepackage{booktabs}
\usepackage{diagbox}
\usepackage{multirow}

\usepackage[normalem]{ulem}

\newcommand{\bra}[1]{\left< #1 \right|} 
\newcommand{\ket}[1]{\left| #1 \right>}

\begin{document}

\bibliographystyle{naturemag_noURL} 

\title{Quantum Fast Hitting on Glued Trees Mapped on a Photonic chip}
\author{Zi-Yu Shi}
\affiliation{State Key Laboratory of Advanced Optical Communication Systems and Networks, School of Physics and Astronomy, Shanghai Jiao Tong University, Shanghai 200240, China}
\affiliation{Synergetic Innovation Center of Quantum Information and Quantum Physics, University of Science and Technology of China, Hefei, Anhui 230026, China}
\author{Hao Tang}
\affiliation{State Key Laboratory of Advanced Optical Communication Systems and Networks, School of Physics and Astronomy, Shanghai Jiao Tong University, Shanghai 200240, China}
\affiliation{Synergetic Innovation Center of Quantum Information and Quantum Physics, University of Science and Technology of China, Hefei, Anhui 230026, China}

\author{Zhen Feng}
\affiliation{State Key Laboratory of Advanced Optical Communication Systems and Networks, School of Physics and Astronomy, Shanghai Jiao Tong University, Shanghai 200240, China}
\affiliation{Synergetic Innovation Center of Quantum Information and Quantum Physics, University of Science and Technology of China, Hefei, Anhui 230026, China}

\author{Yao Wang}
\affiliation{State Key Laboratory of Advanced Optical Communication Systems and Networks, School of Physics and Astronomy, Shanghai Jiao Tong University, Shanghai 200240, China}
\affiliation{Shenzhen Institute for Quantum Science and Engineering and Department of Physics, Southern University of Science and Technology, Shenzhen 518055, China}

\author{Zhan-Ming Li}
\affiliation{State Key Laboratory of Advanced Optical Communication Systems and Networks, School of Physics and Astronomy, Shanghai Jiao Tong University, Shanghai 200240, China}
\affiliation{Synergetic Innovation Center of Quantum Information and Quantum Physics, University of Science and Technology of China, Hefei, Anhui 230026, China}

\author{Zhi-Qiang Jiao}
\affiliation{State Key Laboratory of Advanced Optical Communication Systems and Networks, School of Physics and Astronomy, Shanghai Jiao Tong University, Shanghai 200240, China}
\affiliation{Synergetic Innovation Center of Quantum Information and Quantum Physics, University of Science and Technology of China, Hefei, Anhui 230026, China}

\author{Jun Gao}
\affiliation{State Key Laboratory of Advanced Optical Communication Systems and Networks, School of Physics and Astronomy, Shanghai Jiao Tong University, Shanghai 200240, China}
\affiliation{Shenzhen Institute for Quantum Science and Engineering and Department of Physics, Southern University of Science and Technology, Shenzhen 518055, China}

\author{Yi-Jun Chang}
\affiliation{State Key Laboratory of Advanced Optical Communication Systems and Networks, School of Physics and Astronomy, Shanghai Jiao Tong University, Shanghai 200240, China}
\affiliation{Synergetic Innovation Center of Quantum Information and Quantum Physics, University of Science and Technology of China, Hefei, Anhui 230026, China}

\author{Wen-Hao Zhou}
\affiliation{State Key Laboratory of Advanced Optical Communication Systems and Networks, School of Physics and Astronomy, Shanghai Jiao Tong University, Shanghai 200240, China}
\affiliation{Synergetic Innovation Center of Quantum Information and Quantum Physics, University of Science and Technology of China, Hefei, Anhui 230026, China}

\author{Xian-Min Jin}
\email{xianmin.jin@sjtu.edu.cn} 
\affiliation{State Key Laboratory of Advanced Optical Communication Systems and Networks, School of Physics and Astronomy, Shanghai Jiao Tong University, Shanghai 200240, China}
\affiliation{Synergetic Innovation Center of Quantum Information and Quantum Physics, University of Science and Technology of China, Hefei, Anhui 230026, China}
\affiliation{Shenzhen Institute for Quantum Science and Engineering and Department of Physics, Southern University of Science and Technology, Shenzhen 518055, China}

\begin{abstract}
Hitting the exit node from the entrance node faster on a graph is one of the properties that quantum walk algorithms can take advantage of to outperform classical random walk algorithms. Especially, continuous-time quantum walks on central-random glued binary trees have been investigated in theories extensively for their exponentially faster hitting speed over classical random walks. Here, using heralded single photons to represent quantum walkers and waveguide arrays written by femtosecond laser to simulate the theoretical graph, we are able to demonstrate the hitting efficiency of quantum walks with tree depth as high as 16 layers for the first time. Furthermore, we expand the graph's branching rate from 2 to 5, revealing that quantum walks exhibit more superiority over classical random walks as branching rate increases. Our results may shed light on the physical implementation of quantum walk algorithms as well as quantum computation and quantum simulation.
\end{abstract}

\maketitle 

Due to the unique features like superposition and entanglement, introducing quantum mechanical effects into classical algorithms has shown great speed advantage\cite{montanaro2016quantum,shor1994algorithms,lanyon2007experimental, grover1996fast,grover1997quantum,jones1998implementation,   farhi2001quantum,Aaronson:2004:QLB:1008731.1008735}. Classical random walks (CRWs) on graphs have become a powerful tool for designing classical algorithms. As its counterpart in quantum field, quantum walks\cite{PhysRevA.48.1687}  (QWs) also find wide applications in not only quantum algorithms\cite{ambainis2003quantum, kempe2003quantum,ambainis2007quantum,1751-8121-41-7-075303,childs2004spatial, shenvi2003quantum,tulsi2008faster} but also quantum computation\cite{PhysRevLett.102.180501,Childs791} and quantum simulation\cite{schreiber20122d, peruzzo2010quantum, sansoni2012two, crespi2013anderson}. Fast hitting\cite{kempe2005discrete,PhysRevA.58.915,Childs:2003:EAS:780542.780552} is one of the properties\cite{aharonov2001quantum,moore2002quantum} of QWs on graphs that quantum algorithms can take advange of to demonstrate remarkable quantum speedup, which focuses on the time that a particle needs to reach the exit node from a entrance node with certainty as a function of the size of a graph.

An modified version\cite{Childs:2003:EAS:780542.780552} of the regular glued binary trees \cite{Childs2002} is one of the graphs through which using QWs to traverse is exponentially faster than not only CRWs but also any classical algorithms people can come up with\cite{Childs:2003:EAS:780542.780552}. This structure is constituted by two identical binary trees, and  the depth $n$ of one of the two trees is usually used to describe the size of the whole graph. Each end of the left tree has two branches that are randomly glued to two different ends of the right one, and vice versa (Fig.1a). The left root is the entrance node and the right root is the exit node. We term this structure as central-random glued trees throughout this article. 

\begin{figure*}[ht!]
\includegraphics[width=1.9\columnwidth]{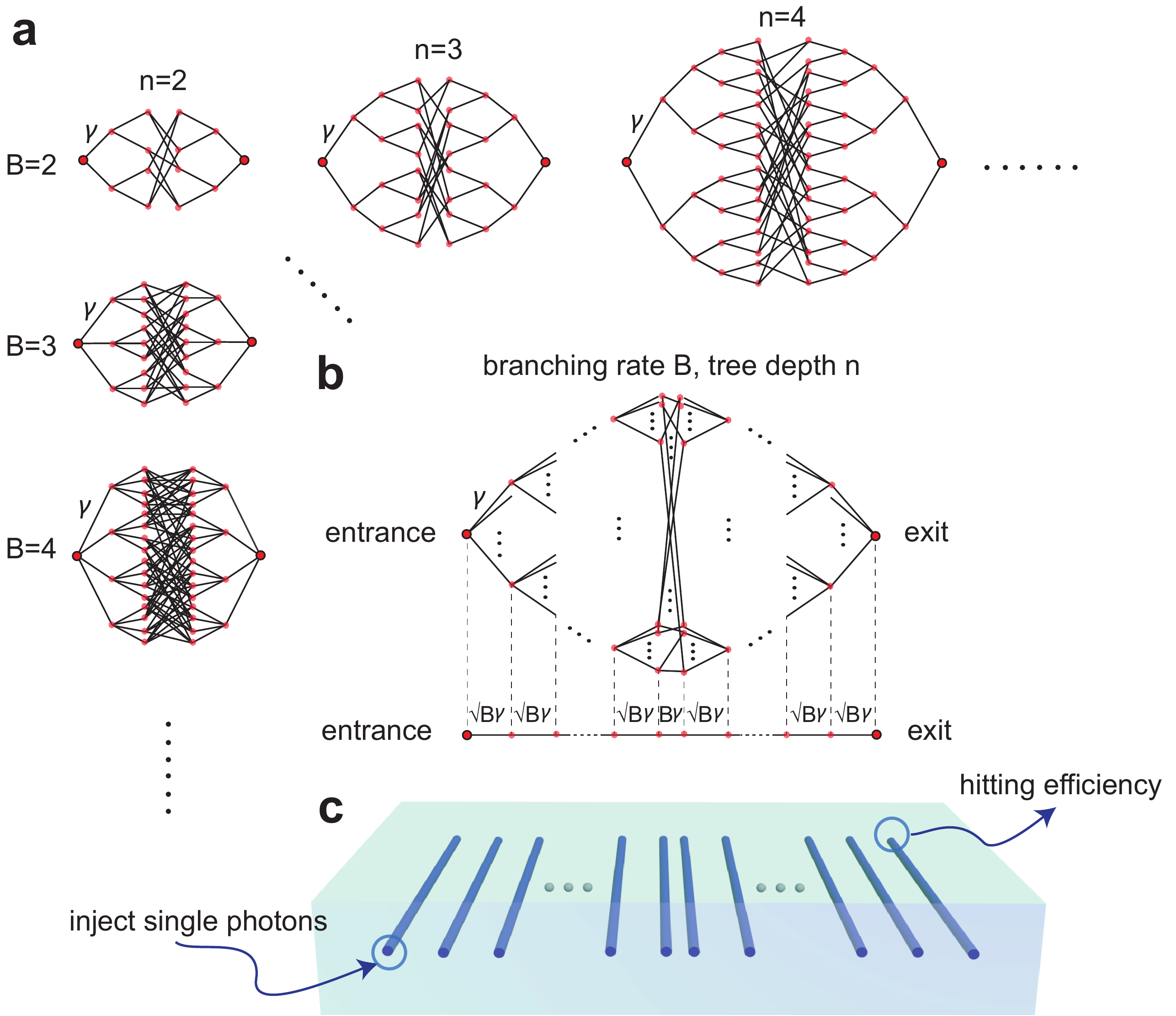}
\caption{\textbf{Schematic diagram of central-random glued trees.} {\bf a.} Central-random glued trees with an increasing $n$ (in row) and $B$ (in column). {\bf b.} Generalized central-random glued trees and its one-dimensional equivalence for QWs. The hopping rate between any adjacent nodes of the two-dimensional graph is $\gamma$, while on the one-dimensional chain, the hopping rate is $\sqrt{B}\gamma$ except for the two adjacent nodes at the center, which becomes $B\gamma$. {\bf c.} Experimental implementation of the mapped one-dimensional chain on a photonic chip using femtosecond laser direct writing technique. The cross section of the waveguide array is consistent with the theoretical one-dimensional chain and the longitudinal direction of the waveguide corresponds to the evolving time. Since the hopping rate of the two adjacent waveguides at the center is greater than others, the spacing between the two waveguides is reduced accordingly.}
\label{fig:QFTConcept}
\end{figure*}

\begin{figure*}[ht!]
\includegraphics[width=1.9\columnwidth]{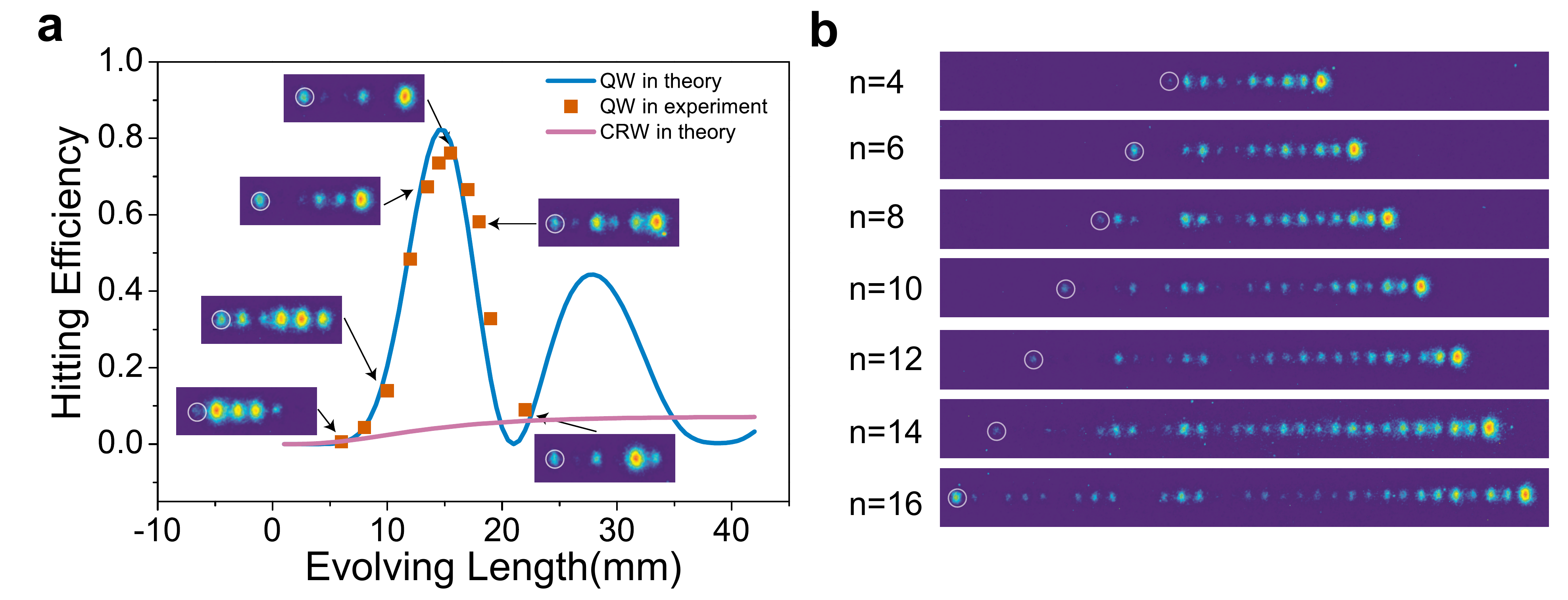}
\caption{\textbf{Spatial photon number distributions and hitting efficiencies for trees at $B=2$.} {\bf a.} Variation of hitting efficiency with evolving length with $n=2$. Several pictures of spatial photon number distributions for samples of different evolving lengths are also shown as insets. The injecting waveguide is marked by a white circle. And corresponding hitting efficiencies of CRWs on central-random glued trees are also plotted. {\bf b.} Spatial photon number distributions that show optimal hitting efficiencies at different $n$.}
\label{fig:strutturaChip}
\end{figure*}

The probability of a CRW hitting the right root\cite{Childs2002}, i.e. the hitting efficiency, is less than $2^{-n}$. Due to the unitarity, the hitting efficiency of QWs always changes with evolving time\cite{Childs2002}, and the first peak occurs in $O(n)$ time and scales as $P_{QW}\sim n^{-{2/3}}$ for large $n$, meaning a QW can almost certainly hit the exit in polynomial time\cite{carneiro2005entanglement,Childs:2003:EAS:780542.780552}. 
What's more, each node on the graph can be extended to $B$ branches rather than just two branches, exhibiting higher complexity (Fig.1b)\cite{carneiro2005entanglement}.
Unfortunately, mapping the full central-random glued trees into a physical system is impossible under existing studies, as the node number grows exponentially with $n$. There has yet been no experimental demonstration of QW's hitting efficiency on this graph, not to mention the expansion to higher branching rates. Even exploiting a deformation\cite{tang2018experimental} or simplified structure\cite{qi2016silicon} is already of great interest.

On the other hand, on account of the coherent superposition of quantum particles as well as the symmetrical shape of this glued trees, the analysis of QW traversing the graph can be reduced to only concerned with the graph's column positions (see Supplementary Note 1). The complex two-dimensional graph can be simplified to a one-dimensional chain with nonuniform hopping rates\cite{Childs:2003:EAS:780542.780552} (Fig.1b). Nonetheless, the physical implementation is still a big challenge compared with those reported uniform structures\cite{perets2008realization, peruzzo2010quantum, Tangeaat3174}, for the relative difference of the chain's hopping rates is the key to simulate different branching rates, which requires high-precision control of the coupling strength between waveguides.

Fortunately, after plenty of time and efforts, we have conquered the difficulties. For the first time, we not only experimentally investigate the variation of QW's optimal hitting efficiency with $n$ ranging from 2 to 16, which demonstrates that a quantum walker can find the exit with polynomially high probability in linear time, but also its variation with $B$ going from 2 to 5, revealing that the branching rate can also contribute to the speed advantage of QWs. Our work will inspire the experimental realization of fast hitting as well as the other properties of QWs\cite{aharonov2001quantum,moore2002quantum,kempe2003quantum,ambainis2003quantum} in more scenarios, further facilitating the physical realization of quantum walk algorithms.

The row of Fig.1a shows the changing of tree depth, while the column exhibits the variation of branching rate. As shown in Fig.1c, in borosilicate chip substrate, we use femtosecond laser direct writing technique\cite{davis1996writing,thomson2011ultrafast,crespi2013integrated,chaboyer2015tunable,feng2016invisibility} to fabricate an array of waveguides, the cross section of which is a strict mapping of the theoretical one-dimensional chain (Fig.1b). Each waveguide represents a node in the theoretical graph and the longitudinal direction of the waveguides is for the evolving process in time. As evolving length $z$ of photons is proportional to evolving time $t$ for $z=vt$, in which $v$ is the propagation speed of photons in waveguides, we use length $z$ to replace time $t$ in the following description.

We prepare and inject single photons into the entrance waveguide, and observe the spatial photon number distribution when the waveguide lengths change. When propagating along the waveguides, a single photon will be coupled evanescently to the other waveguides\cite{Tangeaat3174,Gao:16}. It is worth noting that, since the coupling coefficient decreases exponentially with waveguide spacing, we only consider the coupling effects between the nearest waveguides\cite{Szameit:07,Tangeaat3174}. 
The evolving process of a continuous-time QW can be described by the wavefunction of photons according to$$\ket{\Psi(z)}=e^{-iHz}\ket{\Psi(0)},\eqno{(1)}$$ where $\ket{\Psi(0)}$ is the initial state and $H$ is the Hamiltonian. The equation can be calculated by matrix exponential methods\cite{IZAAC201581}.
The Hamiltonian of photons propagating through a waveguide system can be described by\cite{perets2008realization,peruzzo2010quantum}:
$$
\mathrm{H}=\sum_{i}^{N} \beta_{i} a_{i}^{\dagger} a_{i}+\sum_{i \neq j}^{N} C_{i, j} a_{i}^{\dagger} a_{j}, \eqno{(2)}
$$
where $a_{i}^{\dagger}, a_{i}$ are the creation and annihilation operator for waveguide $i$. $\beta_{i}=\beta$ is the propagation constant. $C_{i, j}$ is the coupling coefficient between waveguide $i$ and $j$, which is mainly determined by waveguide spacing.

 For a central-random glued trees with depth $n$, its one-dimensional equivalence has $2n+2$ nodes and the probabilities of a walker moving right or left (hopping rate) on the chain are same on all the nodes except the two nearest nodes at the center. As Fig.1c shows, on the chain, the hopping rate, which corresponds to the coupling coefficient in experiments, of a quantum walker between the two nearest central nodes is $\sqrt{B}$ times of that between the other nearest nodes, where $\gamma$ is the hopping rate on the trees. What value $\gamma$ is doesn't affect the optimal hitting efficiency, but the ratio of the hopping rate at the center to that at the non-center position, i.e. $\sqrt{B}$, on the chain does, therefore precisely mapping this ratio to waveguide arrays is essential to simulate glued trees with different branching rates. If $j$ represents column positions, then the non-zero elements of the Hamiltonian for the one-dimensional chain are\cite{carneiro2005entanglement}
$$\bra{j}H\ket{j+1} = \left\{\begin{array}{ll}
        \sqrt{B}\gamma & 0\le j<n,\,n<j\le 2n \\
        B\gamma        & j=n\\ \end{array} \right. 
. \eqno{(3)} $$
The rest of the non-zero elements can be deduced by the Hermiticity. The probability distribution of a CRW is uniform and stationary as long as the evolving time is long enough\cite{kempe2003quantum}, whereas the probability distribution of a continuous time QW is always changing.\cite{sanchez2012quantum,tang2018experimental,carneiro2005entanglement}. Several peaks of hitting efficiency which we term as optimal hitting efficiency will appear as the evolving length of photons increases. However, compared with the time cost, the increase in hitting efficiency is negligible and the loss of photons is considerable, so it is desirable to only study the first peak of hitting efficiency and calculate  corresponding evolving length where such first peak occurs, i.e. optimal evolving length. 

\begin{figure*}[ht!]
\includegraphics[width=1.9\columnwidth]{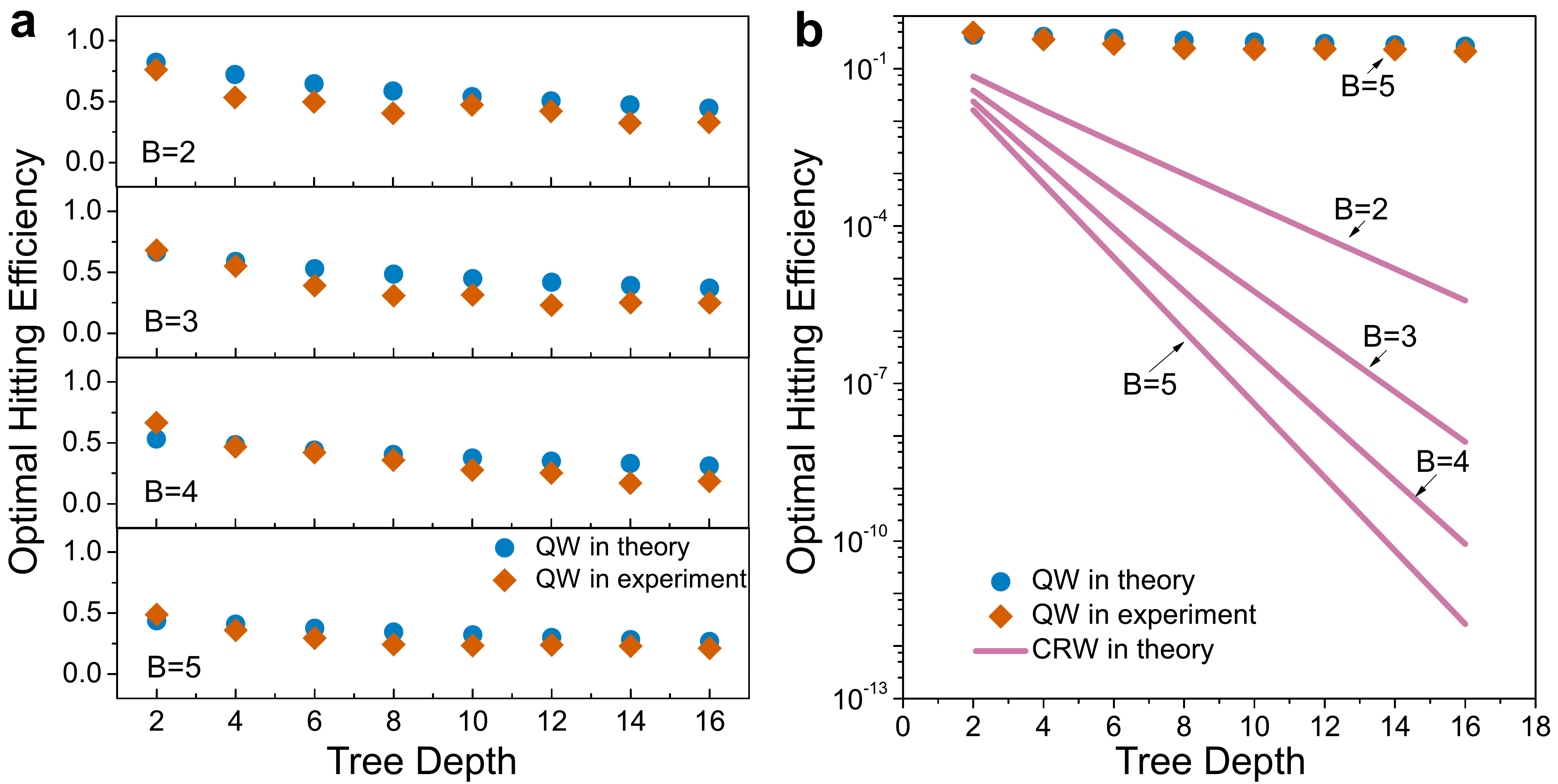}
\caption{\textbf{Experimental optimal hitting efficiencies for different branching rates.} {\bf a.} Variation of optimal hitting efficiencies with n for $B=2, 3, 4, 5$ respectively. As $B$ increases, the optimal hitting efficiencies decrease slightly. {\bf b.} Comparison between CRWs and QWs in a half logarithmic coordinate. The ultimate steady hitting efficiencies of CRWs on central-random blued trees decrease exponentially with $n$, in contrast with the polynomially decreasing trend of QWs. Though data of $B=5$ is the smallest among the four groups, it still outperforms CRWs. } 
\label{fig:apparato}
\end{figure*}

In experiments, on the basis of coupled mode theory, we fabricate a series of waveguide pairs with different waveguide spacings to obtain the function of coupling coefficient with respect to waveguide spacing\cite{Szameit:07}. We then choose suitable coupling coefficients as hopping rates to calculate the variation of hitting efficiency with evolving length. Finally, for graphs of each size, we choose different waveguide lengths as samples to fabricate waveguide arrays with waveguide spacings consistent with the coupling coefficient used in theoretical calculation. Each waveguide is uniform and the distance between any two nearest waveguides is identical except the two waveguides at the center (See Supplementary Note 2 for the fabrication details). With these waveguide arrays in hand, we study the transporting property of QWs by injecting vertically-polarized single photons with a wavelength of 810nm into the entrance waveguide and monitoring the intensity pattern exiting the photonic chip with an intensified charge coupled device (ICCD) camera (See Supplementary Note 3 for details on single-photon imaging). The experimental hitting efficiency is obtained by calculating the ratio of the intensity in the exit waveguide to the total intensity (See Supplementary Note 4 for details on data processing).

For trees with $n$ ranging from 2 to 16, and $B$ ranging from 2 to 5, we first work out the variation of hitting efficiency against evolving length from the experimental patterns. Fig.2a exhibits the results of a 2-layer tree at a branching rate of 2. The experimental optimal hitting efficiency is 0.76 and corresponding evolving length is 15.5mm, agreeing very well with theoretical calculation which suggests an optimal value of 0.82 at 15mm. However, the theoretical hitting efficiency of CRW\cite{Childs:2003:EAS:780542.780552} at 15mm is only 0.044 which is lower than QW for more than one order. Some QW spatial photon number distributions are also added into Fig.2a to visually show the the change of hitting efficiency. The leftmost light spot in each sub-picture corresponds to the entrance waveguide which is marked by a white circle and the rightmost light spot corresponds to the exit waveguide. It can be seen that variation of the ratio of photons reaching the exit waveguide in the spatial photon number distributions is consistent with our calculated experimental curve.

Then, applying the method of finding experimental optimal hitting efficiency as shown in Fig.2a, we investigate the change of optimal hitting efficiency with $n$. Fig.2b shows spatial photon number distribution of optimal hitting efficiency with $n$ going from 2 to 16 at $B=2$. It is obvious that despite the number of waveguides increases with $n$, most of the photons gather at the exit waveguide rather than the other waveguides which represents the positions of the other columns of the tree. All spatial photon number distributions that show optimal hitting efficiency for graphs of different sizes as well as the linearly increasing optimal evolving lengths are presented in supplementary materials (See Supplementary Note 5 and Supplementary Fig.S2-S5). 

The optimal hitting efficiencies calculated from theoretical models and experimental evolution results are plotted in Fig.3a, and the ultimate steady hitting efficiency of CRW when evolving length is long enough is added in Fig.3b as a comparison. We emphasize that this way of reducing to 1D chain is derived for QWs (as shown in Supplementary Note 1), but is not suitable for CRWs, therefore, all theoretical results about CRWs are based on central-random glued trees. Fig.3a exhibits that, for each $B$, QW's optimal hitting efficiency decays polynomially with the increase of $n$. In contrast, for CRW, when evolving length is infinitely long, the hitting efficiency will increase monotonously to an asymptotic value which is the inverse of the number of nodes on the trees\cite{Childs:2003:EAS:780542.780552}, and the scaling can be expressed as $P_{CRW}\sim B^{-{n}}$, meaning its optimal hitting efficiency decays exponentially with $n$\cite{carneiro2005entanglement} (Fig.3b). It can be seen from Fig.3b that though the hitting efficiency of QW at $B=5$ is the smallest among the four kinds of branches, it is still exponentially higher than CRW.

Finally, we investigate the variation of optimal hitting efficiency with $B$. As $B$ increases, the complexity of graphs rises dramatically and the increasing branches at each node provide ever more challenges for CRW particles to choose from. On the other hand, a quantum walker is able to explore the landscape it traverses with superposition, and hence has even more evident superiority over CRW at a higher branching rate. From Fig.3a, the QW's optimal hitting efficiency for $n=16$ at $B=5$ is still more than 50\% of that at $B=2$, while the ratio for the same pair drops to approximately $(2/5)^{16}=4.29 \times 10^{-7}$ for CRW (see Fig.3b), as $P_{CRW}\sim B^{-{n}}$. 

\begin{figure}[!]
	\centering
	\includegraphics[width=1\columnwidth]{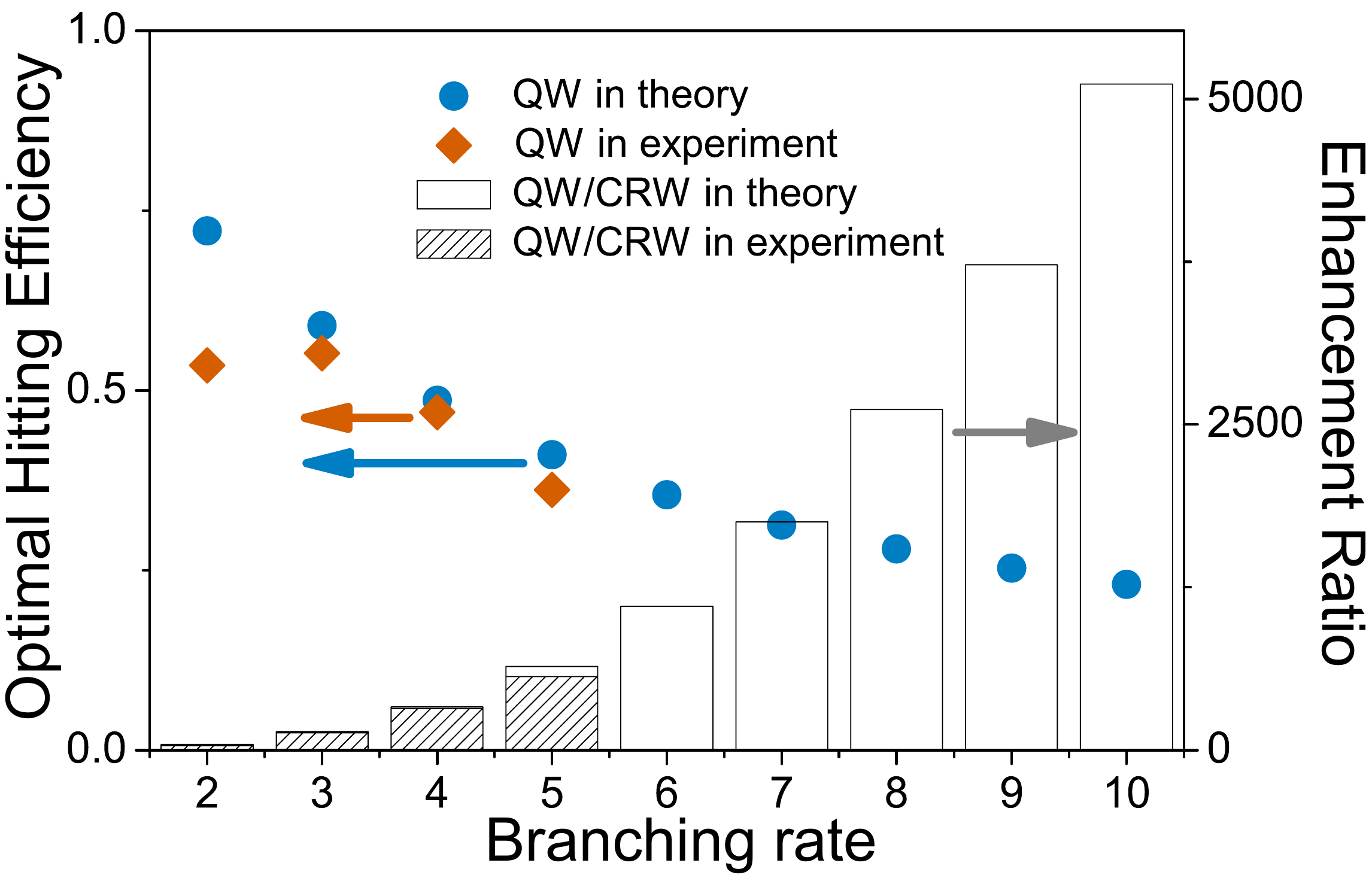}
	\caption{\textbf{Variation of experimental optimal hitting efficiencies with branching rate for $n=4$}. Scatter plots correspond to the left $y$ axis, representing optimal hitting efficiencies of QWs in theory and experiment respectively. Column plots correspond to the right $y$ axis, representing the enhancement rate, i.e. optimal hitting efficiency ratio of QW against CRW. The tree depth remains to be 4 when this measurement is taken.}
	\label{fig:Results4}
\end{figure}

In Fig.4, we plot the QW's optimal hitting efficiency (left $y$ axis) and the enhancement ratio over CRW's optimal hitting efficiency (right $y$ axis) for different branching rates from 2 up to 10. The results all come from 4-layer glued trees. Even for the shallow glued tree without pursuing a large tree depth, the enhancement of optimal hitting efficiency by QW over CRW is impressive at a high branching rate. It has been suggested in theory\cite{carneiro2005entanglement} that when $B$ is considerably large, QW's optimal hitting efficiency scales with $B$ by $P_{QW}\sim 1/B$, and then the enhancement ratio would scale as $r =P_{QW}/P_{CRW}\sim B^{n-1}$ for a certain $n$. We have observed the increasing enhancement ratio for a few branching rates, and provided the first experiment demonstration of central-random glued trees of high branching rates. The experimental results prove that branching rate can be another useful resource besides the more commonly discussed tree depth, to increase the complexity of central-random glued trees for strengthening the quantum superiority.

We also exploit the waveguide arrays used in Fig.4 as samples to measure the second-order anti-correlation parameter\cite{spring2013chip} $\alpha$  of the photons exiting the exit waveguides, which tends to be 0 for ideal single photons and 1 for classical coherent light. For graphs with $B$=2, 3, 4, 5 at $n=4$, the results are shown in TABLE I, demonstrating excellent quantum property, making our setup distinguishable from those using classical coherent light. (See Supplementary note 6 for details)

\begin{table}[htbp]   
\centering
\caption{\textbf{The measured $\alpha$ of the single photons existing from the exit waveguide for graphs with $B$=2, 3, 4, 5 at $n=4$.}}
\label{tab1}
\linespread{1.6}\selectfont
    \begin{tabular}{ p{1.6cm}<{\centering}  p{1.6cm}<{\centering}  p{1.6cm}<{\centering}  p{1.6cm}<{\centering}  p{1.6cm}<{\centering}}    
        \hline            
        $B$ & 2 & 3 & 4 & 5 \\[0.1cm]        
        \hline           
        $\alpha$ & 0.086 & 0.114 & 0.091 & 0.093 \\[0.15cm]        
        errorbar & 0.001 & 0.001 & 0.001 & 0.003 \\[0.15cm]
       
        \hline         
        \end{tabular}
\end{table}


In summary, we demonstrate the fast-hitting property of QWs on central-random glued trees in a single-photon regime. Harnessing femtosecond laser direct writing technique, we precisely control the coupling coefficient between waveguides and fabricate waveguide arrays with high tree depths and branching rates. We employ ultra-low-noise single-photon imaging techniques to measure the probability of heralded single photons reaching the exit waveguides. Using single photons instead of classical coherent light, we manage to observe the accumulated process from the individual single photons to the the eventual spatial photon number distribution, realizing a true sense of quantum-particle experiment instead of the simulation results by laser beams. Besides, using single photons also gives a possible direction for the experimental expansion to a bigger Hilbert space, i.e. increasing the number of particles walking on the graph, which classical coherent light can’t achieve.

We experimentally demonstrate that the optimal hitting efficiency of a quantum walker hitting the exit only decays polynomially as tree depth increases, which is in contrast with CRW's exponentially decreasing trend. We further demonstrate that the enhancement ratio of the QW's optimal hitting efficiency over CRW becomes higher when branching rate increases, suggesting the branching rate as another useful approach besides the tree depth to introduce more evident quantum advantages. 

In fabricating high-branched glued trees, we manage to precisely set coupling coefficients as $\sqrt{B}\gamma$ or $B\gamma$ using the advanced waveguide writing techniques, without which it is impossible to implement such glued trees on photonic chips. This precise writing techniques can be further utilized for richer explorations of on-chip quantum algorithms and quantum simulation. Another inspiring point from this work is the idea of reducing a complex graph to a simpler equivalence that can be mapped in the physical systems. For many complex graphs such as hypercubes and other high-dimensional structures, this may shed light upon the experimental implementation in a highly feasible way. Further, it would also be interesting to map the structure of high branching rates and tree depths to optimization problems in different fields\cite{PhysRevA.58.915}, and bring up quantum advantages by these hitting protocols into real-life applications. 

\section*{Acknowledgements} 
The authors thank Andrew Childs and Jian-Wei Pan for helpful discussions. This work was supported by National Key R\&D Program of China (2017YFA0303700); National Natural Science Foundation of China (NSFC) (61734005, 11761141014, 11690033); Science and Technology Commission of Shanghai Municipality (STCSM) (15QA1402200, 16JC1400405, 17JC1400403); Shanghai Municipal Education Commission (SMEC)(16SG09, 2017-01-07-00-02-E00049); X.-M.J. acknowledges support from the National Young 1000 Talents Plan.

\bigskip

\renewcommand{\bibnumfmt}[1]{#1.}

\clearpage
\newpage

\onecolumngrid
\subsection*{\large Supplemental Information: Quantum Fast Hitting on Glued Trees Mapped on a Photonic chip}
\setcounter{figure}{0}
\setcounter{table}{0}
\setcounter{equation}{0}
\renewcommand{\figurename}{Supplementary Figure}
\renewcommand{\tablename}{Supplementary Table}

\renewcommand{\thetable}{\arabic{table}}
\renewcommand{\theequation}{{S}\arabic{equation}}

\bigskip
\section*{Supplementary Note 1 - Simplification process of the central-random glued trees to a 1D chain}

We will first give the general derivation process [S1], then we will give a simple example to make the general derivation easily understood.

Consider the Hilbert space of the 1D chain spanned by $|\operatorname{col} j\rangle$, which can be represented as the uniform superposition over the nodes in column $j$, i.e.,
$$
|\operatorname{col} j\rangle=\frac{1}{\sqrt{N_{j}}} \sum_{a \in \operatorname{column}j}|a\rangle \eqno{(\rm S1)}
$$
where $|{a}\rangle$ represents the state of a node on the glued trees, and $N_{j}$ is the number of nodes in column $j$.
$$
N_{j}=\left\{\begin{array}{ll}{B^{j}} & {0 \leq j \leq n} \\ {B^{2 n+1-j}} & {n+1 \leq j \leq 2 n+1}\end{array}\right. \eqno{(\rm S2)}
$$

If the adjacency matrix of the central-random glued trees is represented by $A$, in which, if two nodes $i$ and $j$ are connected, $A_{i, j}=1$, otherwise, $A_{i, j}=0$, then we have the following derivation process. We assume the hopping rate $\gamma$ of central-random glued trees is 1 in the whole derivation process.
For any $0<j<n$, we have
\begin{align}
 A|\operatorname{col} j\rangle
&=\frac{1}{\sqrt{N_{j}}} \sum_{a \in \operatorname{column} j} A|a\rangle\tag{\rm S3} \\
\nonumber &=\frac{1}{\sqrt{N_{j}}}\left(B \sum_{a \in \operatorname{column} j-1}|a\rangle+\sum_{a \in \operatorname{column} j+1}|a\rangle\right)\\
\nonumber & =\frac{1}{\sqrt{N_{j}}}\left(B \sqrt{N_{j-1}}|\operatorname{col} j-1\rangle+\sqrt{N_{j+1}}|\operatorname{col} j+1\rangle\right)\\
\nonumber & =\sqrt{B}(|\operatorname{col} j-1\rangle+|\operatorname{col} j+1\rangle). 
\end{align}

In a similar way, for any $n+1<j<2n+1$, we have
\begin{align}
 A|\operatorname{col} j\rangle
&=\frac{1}{\sqrt{N_{j}}}\left(\sum_{a \in \operatorname{column} j-1}|a\rangle+ B\sum_{a \in \operatorname{column} j+1}|a\rangle\right)\tag{\rm S4} \\
\nonumber & =\frac{1}{\sqrt{N_{j}}}\left( \sqrt{N_{j-1}}|\operatorname{col} j-1\rangle+B\sqrt{N_{j+1}}|\operatorname{col} j+1\rangle\right)\\
\nonumber & =\sqrt{B}(|\operatorname{col} j-1\rangle+|\operatorname{col} j+1\rangle).
\end{align}

We can see that the hopping rates on the chain which correspond to the mapping of the left tree and the right tree are uniform, and are $\sqrt{B}$ times of that on the original graph. 

However, when comes to the mapping of the random-glued part of the glued trees, the results are different. The random-glued part has to satisfy the condition that each node in column $n$ should be connected to $B$ different nodes in column $n+1$, and vice versa. Then, if $j=n$, we have
\begin{align}
 A|\operatorname{col} n\rangle
&=\frac{1}{\sqrt{N_{n}}}\left(B\sum_{a \in \operatorname{column} n-1}|a\rangle+B \sum_{a \in \operatorname{column} n+1}|a\rangle\right)\tag{\rm S5} \\
\nonumber & =\frac{1}{\sqrt{N_{n}}}\left( B\sqrt{N_{n-1}}|\operatorname{col} n-1\rangle+B\sqrt{N_{n+1}}|\operatorname{col} n+1\rangle\right)\\
\nonumber & =\sqrt{B}|\operatorname{col} n-1\rangle+B|\operatorname{col} n+1\rangle.
\end{align}
Similarly, 
\begin{align}
A|\operatorname{col} n+1\rangle
&=\frac{1}{\sqrt{N_{n+1}}}\left(B\sum_{a \in \operatorname{column} n}|a\rangle+B \sum_{a \in \operatorname{column} n+2}|a\rangle\right)\tag{\rm S6} \\
\nonumber & =\frac{1}{\sqrt{N_{n+1}}}\left(B \sqrt{N_{n}}|\operatorname{col} n\rangle+B\sqrt{N_{n+2}}|\operatorname{col} n+2\rangle\right)\\
\nonumber & =B|\operatorname{col} n\rangle+\sqrt{B}|\operatorname{col} n+2\rangle.
\end{align}

We can see that the hopping rate on the chain corresponding to the random-glued part  is $B$ times of the hopping rate on the glued trees, different from the rest of the hopping rates. It is obvious that what exactly the random-glued part is doesn't affect the geometry of the 1D chain. Next, we will use a simple example to illustrate the above derivation process.

\begin{figure*}[ht!]
\includegraphics[width=0.6\textwidth]{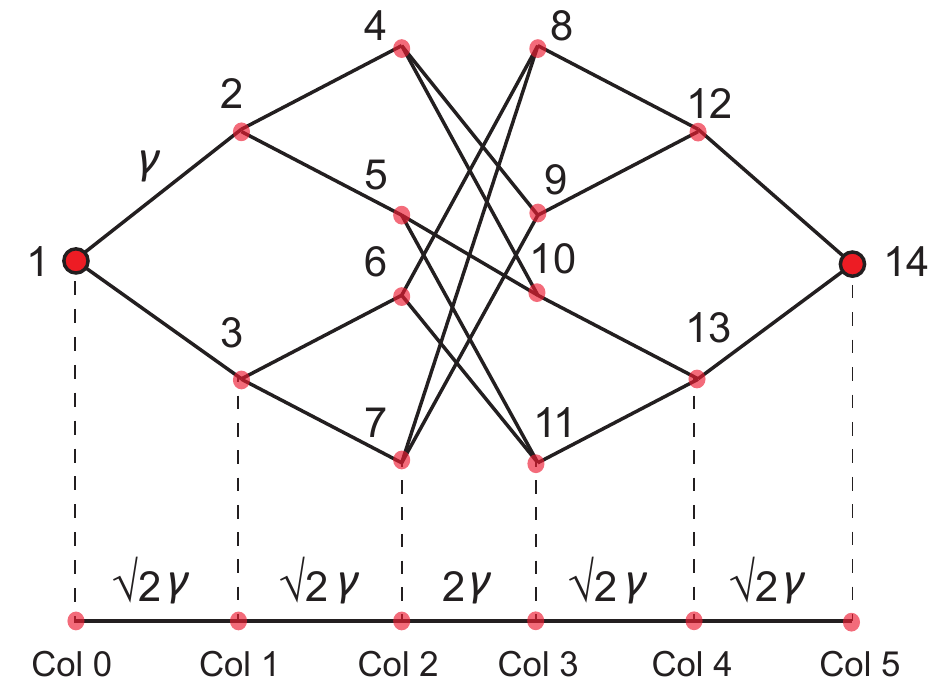}
\caption{\textbf{Schematic diagram of central-random glued trees with B=2, n=2}.  }
\end{figure*}

If $|\operatorname{col} 0\rangle$, $|\operatorname{col} 1\rangle$, $|\operatorname{col} 2\rangle, |\operatorname{col} 3\rangle$ can be represented as
$$
|\operatorname{col} 0\rangle=|1\rangle,\eqno{(\rm S7)}
$$
$$
|\operatorname{col} 1\rangle=\frac{1}{\sqrt{2}}(|2\rangle+|3\rangle),\eqno{(\rm S8)}
$$
$$
|\operatorname{col} 2\rangle=\frac{1}{\sqrt{4}}(|4\rangle+|5\rangle+|6\rangle+|7\rangle),\eqno{(\rm S9)}
$$
$$
|\operatorname{col} 3\rangle=\frac{1}{\sqrt{4}}(|8\rangle+|9\rangle+|10\rangle+|11\rangle),\eqno{(\rm S10)}
$$
then,
\begin{align}
 A|\operatorname{col} 1\rangle &=\frac{1}{\sqrt{2}} A(|2\rangle+|3\rangle)\tag{\rm S11}\\
\nonumber & =\frac{1}{\sqrt{2}}(|1\rangle+|4\rangle+|5\rangle+|1\rangle+|6\rangle+|7\rangle\rangle\\
\nonumber & =\frac{1}{\sqrt{2}}(2|\operatorname{col} 0\rangle+\sqrt{4}|\operatorname{col} 2\rangle)\\
\nonumber & =\sqrt{2}(|\operatorname{col} 0\rangle+|\operatorname{col} 2\rangle).
\end{align}
\begin{align}
A|\operatorname{col} 2\rangle &=\frac{1}{\sqrt{4}} A(|4\rangle+|5\rangle+|6\rangle+|7\rangle)\tag{\rm S12}\\
\nonumber & =\frac{1}{\sqrt{4}}(|2\rangle+|9\rangle+|10\rangle+|2\rangle+|10\rangle+|11\rangle+|3\rangle+|8\rangle+|11\rangle+|3\rangle+|8\rangle+|9\rangle\rangle\\
\nonumber & =\frac{1}{\sqrt{4}}(2\sqrt{2}|\operatorname{col} 1\rangle+2\sqrt{4}|\operatorname{col} 3\rangle)\\
\nonumber & =\sqrt{2}|\operatorname{col} 1\rangle+2|\operatorname{col} 3\rangle.
\end{align}
The hopping rates between $|\operatorname{col} 0\rangle$ and $|\operatorname{col} 1\rangle$, $|\operatorname{col} 1\rangle$ and $|\operatorname{col} 2\rangle$, $|\operatorname{col} 2\rangle$ and $|\operatorname{col} 3\rangle$ on the chain are $\sqrt{2}$, $\sqrt{2}$, $2$ respectively. It is obvious that, as long as the random-glued part satisfies that each node in column 2 is connected to 2 different nodes in column 3, and vice versa for column 3, the specific connection way doesn't affect the structure of the 1D chain. The hopping rates between the remaining nodes on the chain can be derived in a similar way.

\section*{Supplementary Note 2 - Fabrication of the waveguide arrays}
In the femtosecond laser direct writing process, the writing laser with a wavelength of 513nm is up-converted from a femtosecond laser with a wavelength of 1026nm, a pulse duration of 290~fs, and a repetition rate of 1~MHz, and is firstly sent through a spatial light modulation (SLM) and then focused on a borosilicate chip substrate through a 50$\times$ objective lens. The waveguide is written at a depth of 380um with a velocity of 10mm/s. To insure the uniformity of the waveguide arrays, power and depth compensation are also used.\\

\section*{Supplementary Note 3 - Single-photon imaging of spatial photon number distributions}
To carry out the experiment in a quantum regime, we use heralded single photons as the photon source (FIG. S6). The 810nm photon pairs are generated via a type-II spontaneous parametric down conversion (SPDC) process, pumped by an ultraviolet laser with a wavelength of 405nm in a PPKTP crystal. The generated photon pairs are sent through a long-pass filter and then separated into horizontal and vertical component by utilizing a polarized beam splitter(PBS). The vertically-polarized photons are injected into the waveguide arrays, playing the role of quantum walkers, while the horizontally-polarized photons are used to give the ICCD camera a photographing command via a single-photon detector (APD). The photographing commands triggered by the horizontally-polarized photons are used to avoid the impact of thermal light statistics.\\

\section*{Supplementary Note 4 - Calculation of the hitting efficiency}
By photographing the spatial photon number distributions of single photons via an ICCD, we obtain the picture as well as a corresponding ASCII file which records the photon intensity of each pixel. We first find the center pixel coordinate for each waveguide shown in the picture, and then measure the radius in terms of pixel counts for the light spot in each waveguide. By summing up the intensity value within the radius and normalizing them, we get the probabilities of photons distributed at each waveguide.

\section*{Supplementary Note 5 - Spatial photon number distributions and scalings of optimal evolving length}

The experimental results for trees with $B=$2, 3, 4, 5 are presented in FIG. S2-S5 respectively. Each figure shows spatial photon number distributions of optimal hitting efficiencies and scalings of corresponding optimal evolving lengths with $n$ ranging from 2 to 16. Optimal evolving length refers to the evolving length at which optimal hitting efficiency occurs. Note that hopping rate $\gamma$ between two adjacent nodes on central-random glued trees will influence the optimal evolving length, hence we set the same hopping rate for all the four scaling plots shown in FIG. S2-S5. The linear scaling is clearly suggested by theoretical calculation and is well agreed by experimental results. For samples at the same $B$, though optimal hitting efficiency decreases slightly with the increasing number of waveguides as $n$ increases, most of the photons would gather at the exit waveguide when optimal hitting efficiency occurs. Comparing spatial photon number distributions of the same $n$ in different $B$s, there is an overall drop of the light intensity at the exit waveguide from $B=2$ to $B=5$. This leads to the slight decrease of optimal hitting efficiency as $B$ increases, and is consistent with theoretical studies [S2].

\begin{figure*}[ht!]
\includegraphics[width=1\textwidth]{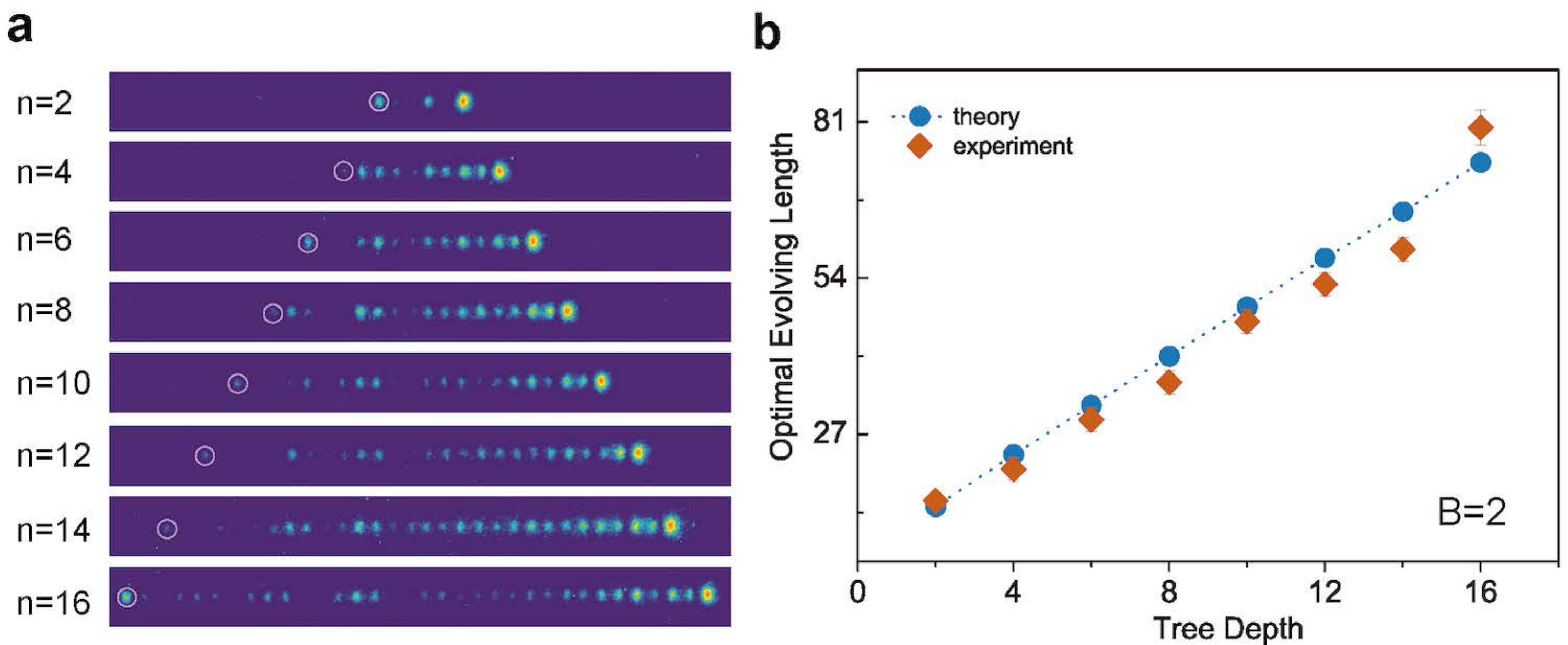}
\caption{\textbf{Spatial photon number distributions and scalings of optimal evolving lengths for samples at $B=2$}. {\bf a,} Spatial photon number distributions of optimal hitting efficiency from 2-layer to 16-layer at $B=2$. The injecting waveguide is marked by a white circle. {\bf b,} Variation of optimal evolving lengths with $n$ ranging from 2 to 16 in theory and experiment respectively. Error bars for the experimental optimal evolving length are the intervals of evolving length values between two adjacent samples at the same $n$ and $B$ used in experiments. Since evolving length values are discrete in experiment, the optimal evolving length may lie between two adjacent length values. The error bar descriptions also apply to FIG. S2, S3 and S4.}
\label{fig:Results4}
\end{figure*}

\begin{figure*}[ht!]
\includegraphics[width=1\textwidth]{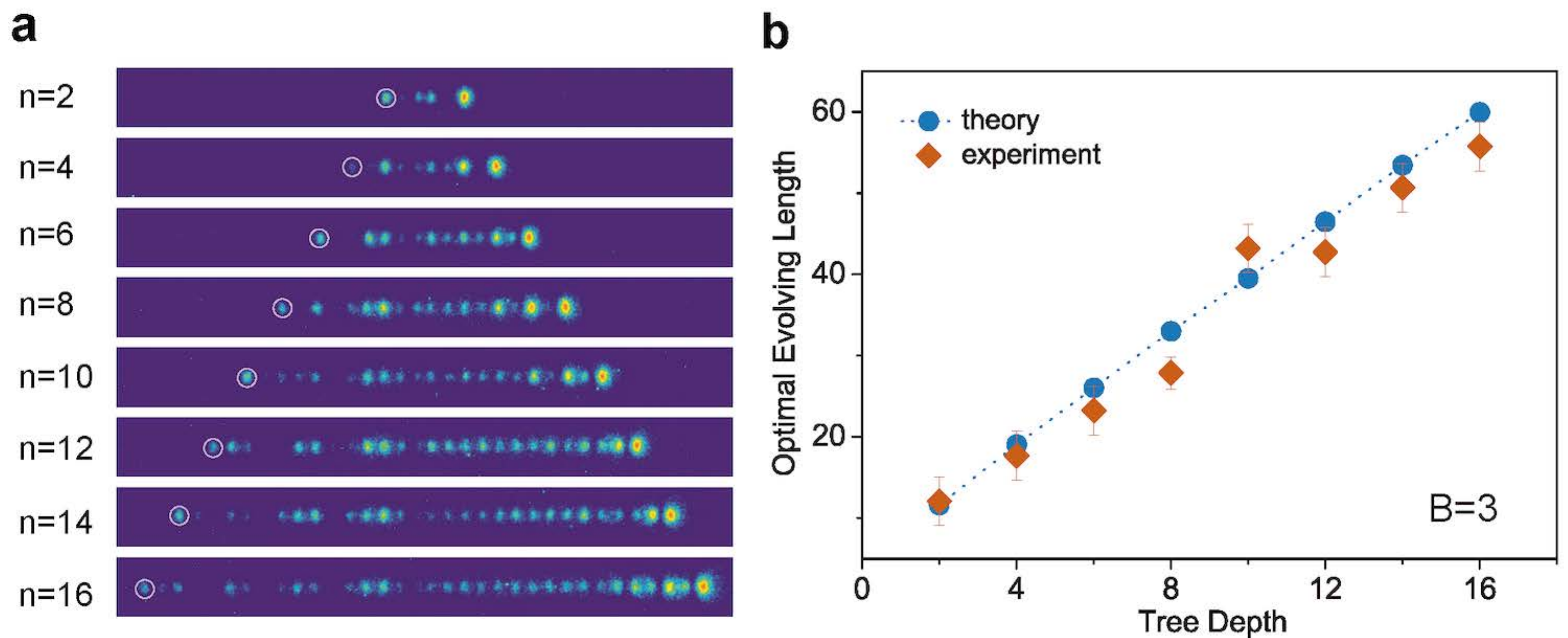}
\caption{\textbf{Spatial photon number distributions and scalings of optimal evolving lengths for samples at $B=3$}.  {\bf a,} Spatial photon number distributions of optimal hitting efficiency from 2-layer to 16-layer at $B=3$. The injecting waveguide is marked by a white circle. {\bf b,} Variation of optimal evolving lengths with $n$ ranging from 2 to 16 in theory and experiment respectively.}
\label{fig:Results4}
\end{figure*}

\begin{figure*}[ht!]
\includegraphics[width=1\textwidth]{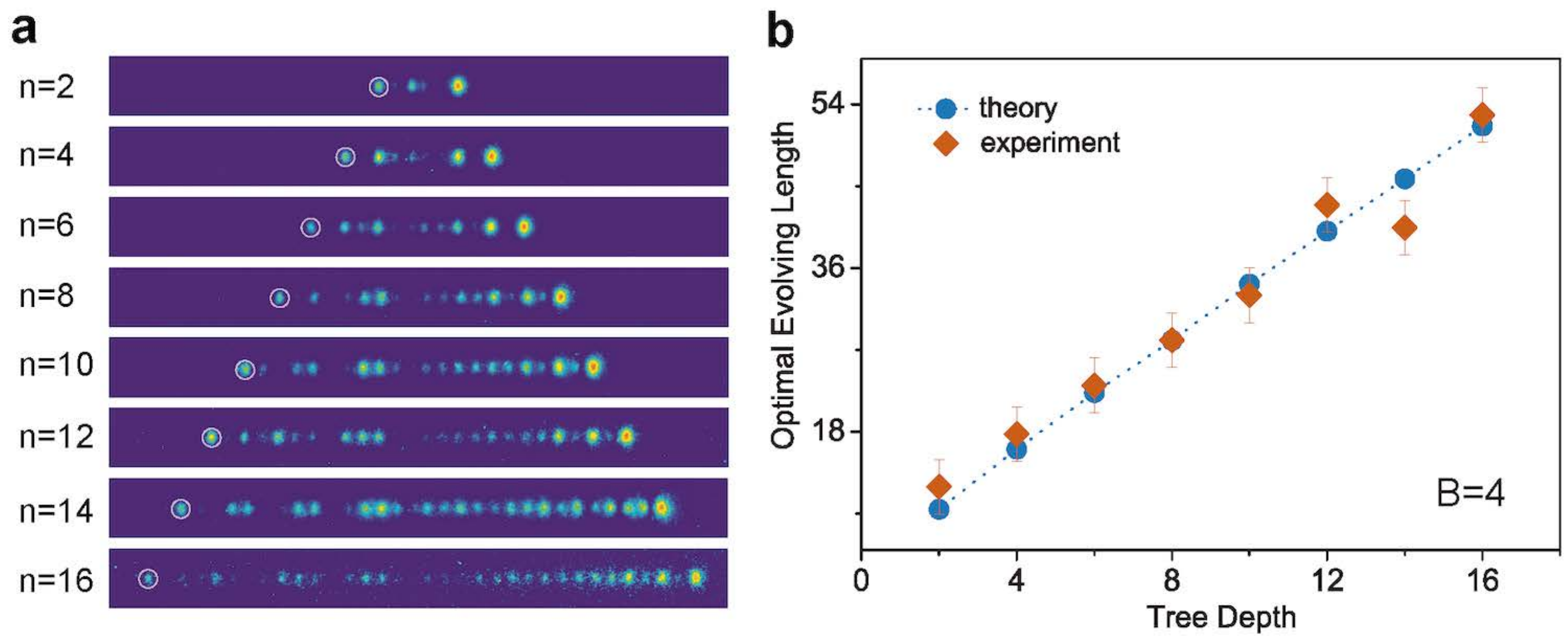}
\caption{\textbf{Spatial photon number distributions and scalings of optimal evolving length for samples at $B=4$}. {\bf a,} Spatial photon number distributions of optimal hitting efficiency from 2-layer to 16-layer at at $B=4$. The injecting waveguide is marked by a white circle. {\bf b,} Variation of optimal evolving lengths with $n$ ranging from 2 to 16 in theory and experiment respectively.}
\label{fig:Results4}
\end{figure*}


\begin{figure*}[h!]
\includegraphics[width=1\textwidth]{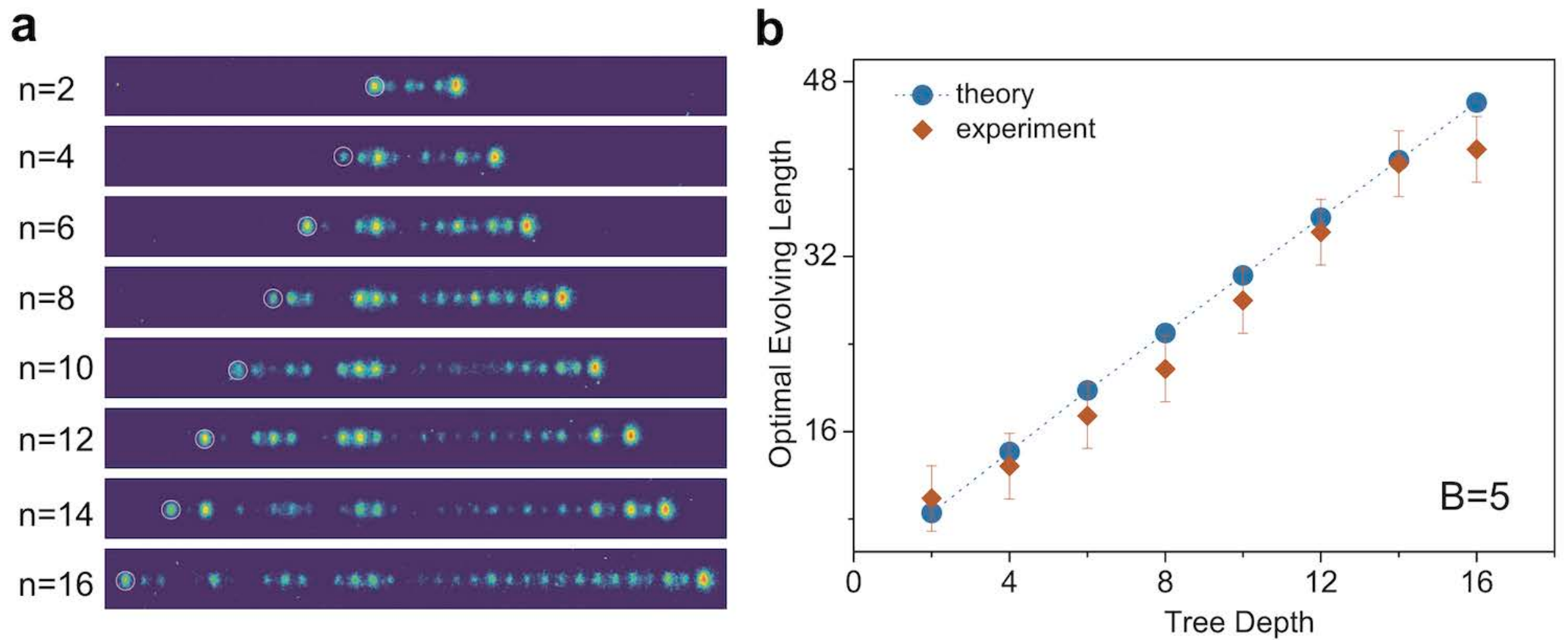}
\caption{\textbf{Spatial photon number distributions and scalings of optimal evolving length for samples at $B=5$}. {\bf a,} Spatial photon number distributions of optimal hitting efficiency from 2-layer to 16-layer at $B=5$. The injecting waveguide is marked by a white circle. {\bf b,} Variation of optimal evolving lengths with $n$ ranging from 2 to 16 in theory and experiment respectively.}
\label{fig:Results4}
\end{figure*}

\clearpage

\section*{Supplementary Note 6 - Measurements of second-order anti-correlation parameter $\alpha$}

\begin{figure*}[ht!]
\includegraphics[width=1\textwidth]{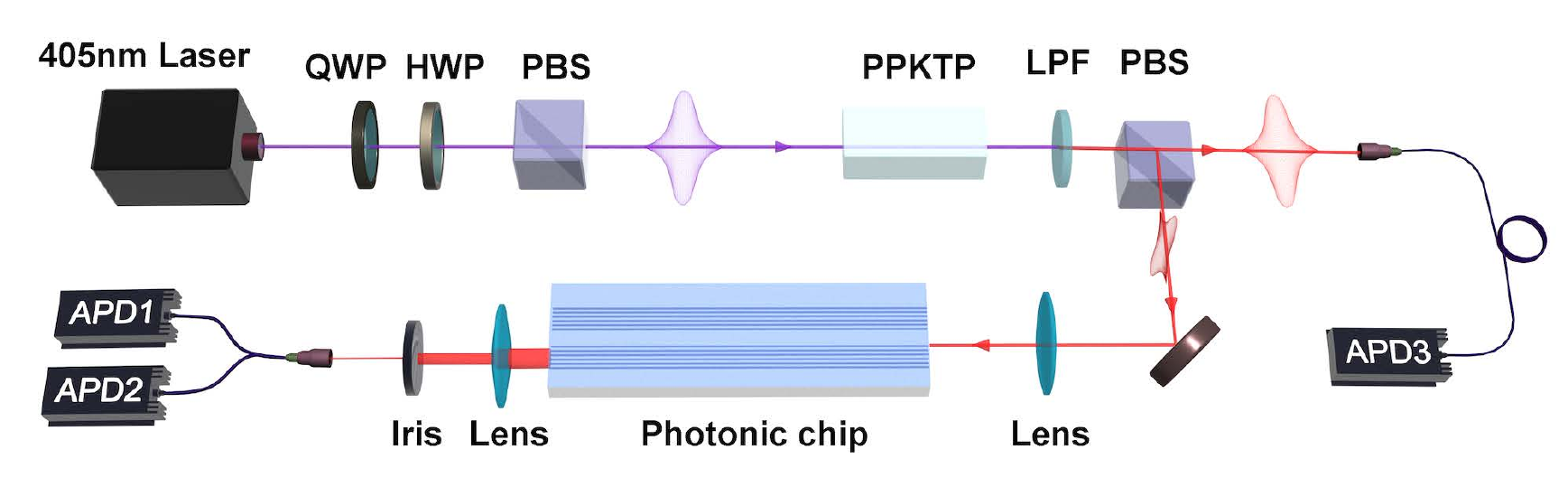}
\caption{\textbf{Setup of measuring $\alpha$}. A 405nm laser pumping a PPKTP crystal can generate 810nm correlated photon pairs via the type-II SPDC process. A long pass filter (LPF) is inserted to block the pump laser. Then the photon pairs pass through a PBS and are separated into vertically-polarized photons and horizontally-polarized photons. The horizontally-polarized photons play the rule of trigger signal and are detected by an avalanched photo diode (APD3); the vertically-polarized photons are injected into the photonic chip. Then, an iris is used to filter out the photons coming from the exit waveguide. Finally, the out-coming photons are coupled into a fiber beam splitter and detected by two separate APDs (APD1 and APD2). A photon coincidence counter module (not shown in the picture) is utilized to record the coincidence events. This setup can be switched into the single-photon imaging of spatial photon number distribution by replacing APD1, APD2 and fiber beam splitter with an ICCD camera. QWP, quarter-wave plate; HWP, half-wave plate; PBS, polarized beam splitter; PPKTP, periodically poled KTP crystal; LPF, long-pass filter; APD, avalanched photo diode.}
\label{fig:Results6}
\end{figure*}

The second-order anti-correlation parameter $\alpha$, which tends to be 0 for ideal single photon and 1 for classical coherent light, can be described as [S3]
$$
\alpha=\frac{{N}_{3} {N}_{123}}{{N}_{13} {N}_{23}},\eqno{(\rm S13)}
$$where $N_3$ represents the photon number of trigger signal; $N_{23}$($N_{13}$), $N_{123}$ represent the number of two- and three-fold coincidence detection events.

As shown in FIG. S6, one photon of the single photon pairs that generated in the SPDC single photon source is used as the trigger signal, and another is injected into the waveguide arrays. The number of trigger signal can be detected by an avalanched photo diode (APD3), through which we can obtain $N_3$. As for the photons being injected into the waveguide arrays, when they have exited the photonic chip, those coming from the exit waveguide are filtered out by an iris inserted after the chip, then the out-coming photons are coupled into a fiber beam splitter and detected by two separate APDs (APD1 and APD2), hence we can obtain the rest of the coincidence detection events. A photon coincidence counter module (not shown in the picture) is utilized to record the coincidence events.

\section*{Supplementary Reference}


\noindent [S1] Retrieved from: \url{https://www.math.uwaterloo.ca/~amchilds/teaching/w08/l13.pdf}\\
\noindent [S2] Carneiro, I., \emph{et al.}  Entanglement in coined quantum walks on regular graphs. \emph{New J. Phys.} {\bf 7}, 156 (2005)\\
\noindent [S3] Spring, J. B., \emph{et al.}  On-chip low loss heralded source of pure single photons. \emph{Opt. Express} {\bf 21}, 13522-13532 (2013)\\

\end{document}